\documentclass[preprint,preprintnumbers,amsmath,amssymb]{revtex4}
\usepackage{epsfig}
\usepackage{graphicx}
\usepackage{dcolumn}
\usepackage{bm}
\usepackage{threeparttable}

\def\beq{\begin{equation}}
\def\eeq{\end{equation}}
\def\eeqn{\end{equation}}
\newcommand\iden{\leavevmode\hbox{\small1\normalsize\kern-.33em1}}

\newcommand{\bea} {\begin{eqnarray}}
\newcommand{\eea} {\end{eqnarray}}

\let\jnfont=\rm
\def\NPB#1,{{\jnfont Nucl.\ Phys.\ B }{\bf #1},}
\def\PLB#1,{{\jnfont Phys.\ Lett.\ B }{\bf #1},}
\def\EPJC#1,{{\jnfont Eur.\ Phys.\ Jour.\ C }{\bf #1},}
\def\PRD#1,{{\jnfont Phys.\ Rev.\ D }{\bf #1},}
\def\PRL#1,{{\jnfont Phys.\ Rev.\ Lett.\ }{\bf #1},}
\def\MPLA#1,{{\jnfont Mod.\ Phys.\ Lett.\ A }{\bf #1},}
\def\JPG#1,{{\jnfont J.\ Phys.\ G }{\bf #1},}
\def\CTP#1,{{\jnfont Commun.\ Theor.\ Phys.\ }{\bf #1},}
\def\JHEP#1,{{\jnfont JHEP \ }{\bf #1},}
\def\NPPS#1,{{\jnfont Nucl.\ Phys.\ Proc.\ Suppl.\ }{\bf #1},}
\def\CPC#1,{{\jnfont Comput.\ Phys.\ Commun.\ }{\bf #1},}
\def\CPL#1,{{\jnfont Chin.\ Phys.\ Lett. }{\bf #1},}
\def\APPB#1,{{\jnfont Acta\ Phys.\ Polon.\ B }{\bf #1},}
\def\PR#1,{{\jnfont Phys.\ Rept.\  }{\bf #1},}
\def\CHC#1,{{\jnfont Chin.\ Phys.\ C }{\bf #1},}

\def\lsim{\raise0.3ex\hbox{$<$\kern-0.75em\raise-1.1ex\hbox{$\sim$}}}
\def\gsim{\raise0.3ex\hbox{$>$\kern-0.75em\raise-1.1ex\hbox{$\sim$}}}

\begin{document}

\title{\ \\[10mm] muon $g-2$ anomaly and $\mu$-$\tau$-philic Higgs doublet with a light CP-even component}

\author{Hong-Xin Wang$^{1}$, Lei Wang$^{1}$, Yang Zhang$^{2}$}
 \affiliation{$^1$ Department of Physics, Yantai University, Yantai
264005, P. R. China\\
$^2$ School of Physics and Microelectronics, Zhengzhou University, ZhengZhou 450001, P. R. China
}


\begin{abstract}
We examine the possibilities of accommodating the
muon $g-2$ anomaly reported by Fermilab in the 2HDM with a discrete $Z_4$ symmetry in
which an inert Higgs doublet field ($H,~A,~H^\pm$) has the lepton flavor violation $\mu$-$\tau$ interactions. 
We study the case of light $H$ (5 GeV $<m_H<$ 115 GeV) and assume the Yukawa matrices to be real and symmetrical.
Considering relevant theoretical and experimental constraints, especially for the
multi-lepton searches at the LHC, we find the muon $g-2$ anomaly can be explained within $2\sigma$ range
in the region of 5 GeV $<m_H<20$ GeV, 130 GeV $< m_A~(m_{H^\pm})<$ 610 GeV, and 0.005 $<\rho<$ 0.014. Meanwhile,
the $\chi^2_\tau$ fitting the data of lepton flavour universality in the $\tau$ decays
approaches to the SM prediction.
\end{abstract}

\maketitle

\section{Introduction}
The Fermilab collobartion presented the new result for
muon anomalous magnetic moment $g-2$ which now, combined with the
data of the E821 \cite{mug2-exp}, amounts to \cite{fermig2}
\bea
\Delta a_\mu=a_\mu^{exp}-a_\mu^{SM}=(251\pm59)\times10^{-11}.
\eea 
The experimental value has an
approximate $4.2\sigma$ discrepancy from the SM prediction.

Two-Higgs-doublet model (2HDM) is a simple extension of SM by adding another Higgs doublet field.
The muon $g-2$ anomaly can be simply explained in the lepton-specific 2HDM \cite{mu2h1,mu2h2,mu2h3,mu2h5,mu2h9,mu2h11,mu2h14,
mu2h15,mu2h16,tavv-1,crivellin,tavv-2,mu2h19} and aligned 2HDM \cite{mua2h1,mua2h2,mua2h3,mua2h4,mua2h5,mua2h6,mua2h7,mua2h8,mua2h9,mua2h10}.
However, the tree-level diagram mediated by the charged Higgs gives negative contribution
to the decay $\tau\to \mu\nu\bar{\nu}$, which will raise the deviation of the LFU in
$\tau$ decays \cite{tavv-1,crivellin,tavv-2}. 
Besides, a scalar with the $\mu$-$\tau$ LFV interactions can accommodate the muon $g-2$ anomaly by
the contribution of one-loop diagrams \cite{0207302,10010434,150207824,151108544,151108880,1610.06587,160604408,190410908,1907.09845,1908.09845,lfv11,lfv12,lfv13}.
Meanwhile, the extra Higgs doublet with the $\mu$-$\tau$ LFV interactions can alleviate the discrepancy of LFU in $\tau$ decays \cite{190410908}.
In this paper, we consider relevant theoretical and experimental constraints, including the LUF in the $\tau$ decays and 
 multi-lepton event searches at the LHC, and examine the possibilities of explaining
 the muon $g-2$ anomaly reported by Fermilab in the 2HDM with a discrete $Z_4$ symmetry in
which an inert Higgs doublet field ($H,~A,~H^\pm$) has the lepton flavor violation $\mu$-$\tau$ interactions. 
Ref. \cite{1908.09845} applied the model to discuss the E821 result of muon $g-2$ anomaly combining the LUF in the $\tau$ decays and 
 multi-lepton event searches at the LHC, and examined the case of $m_H>$ 200 GeV.
Different from the Ref. \cite{1908.09845}, we study the case of light $H$ (5 GeV $<m_H<$ 115 GeV) in the paper.

Our work is organized as follows. In Sec. II we introduce the
model briefly. In Sec. III we discuss the muon $g-2$, the LUF in $\tau$ decays, the exclusion limits of multi-lepton event searches at the LHC, 
and other relevant constraints. In Sec. IV, we show the allowed and excluded paramter space.
Finally, we give our conclusion in Sec. V.

\section{The 2HDM with $\mu$-$\tau$-philic Higgs doublet}
The SM is extended by adding an inert Higgs doublet $\Phi_2$ 
under an abelian discrete $Z_4$ symmetry, and the $Z_4$ charge assignment 
is shown in Table I \cite{190410908}.
\begin{table}
\caption{The $Z_4$ charge assignment.}
\label{tab:matter}
\begin{tabular}{|c||c|c|c||c|c|c||c|c|c||c|c|}
\hline
&~~$Q_L^{i}$~~&~~$U_R^i~~$&~~$D_R^i$~~&~~$L_L^e$~~&~~$L_L^\mu$~~&~~$L_L^\tau$~~&~~$e_R$~~&~~$\mu_R$~~&~~$\tau_R$~~&~~$\Phi_1$~~&~~$\Phi_2$~~\\ \hline
    ~~Z$_4$~~& 1         & 1       & 1       & 1          & $i$             & $-i$          & 1     & $i$        & $-i$     & $1$  & -1     \\ \hline
\end{tabular}
\end{table}
The scalar potential is
given as
\begin{eqnarray} \label{V2HDM} \mathrm{V} &=& Y_1
(\Phi_1^{\dagger} \Phi_1) + Y_2 (\Phi_2^{\dagger}
\Phi_2)+ \frac{\lambda_1}{2}  (\Phi_1^{\dagger} \Phi_1)^2 +
\frac{\lambda_2}{2} (\Phi_2^{\dagger} \Phi_2)^2  \nonumber \\
&&+ \lambda_3
(\Phi_1^{\dagger} \Phi_1)(\Phi_2^{\dagger} \Phi_2) + \lambda_4
(\Phi_1^{\dagger}
\Phi_2)(\Phi_2^{\dagger} \Phi_1)\nonumber \\
&&+ \left[\frac{\lambda_5}{2} (\Phi_1^{\dagger} \Phi_2)^2 + \rm
h.c.\right].
\end{eqnarray}
We discuss the CP-conserving scenario in which all
$\lambda_i$ are real. The two complex
scalar doublets can be given as
\begin{equation} \label{field}
\Phi_1=\left(\begin{array}{c} G^+ \\
\frac{1}{\sqrt{2}}\,(v+h+iG^0)
\end{array}\right)\,, \ \ \
\Phi_2=\left(\begin{array}{c} H^+ \\
\frac{1}{\sqrt{2}}\,(H+iA)
\end{array}\right). \nonumber
\end{equation}
The $\Phi_1$ field has the vacuum expectation value (VEV) $v$=246
GeV, and the VEV of $\Phi_2$ field is zero. The $Y_1$ is determined by requiring the scalar
potential minimization condition.
\beq
Y_1=-\frac{1}{2}\lambda_1 v^2.
\eeq

The Nambu-Goldstone bosons $G^0$ and $G^+$ are eaten by the gauge bosons. 
The $H^+$ and $A$ are the mass eigenstates of the charged Higgs boson and
CP-odd Higgs boson.  
 Their masses are given as
\beq \label{masshp}
 m_{H^\pm}^2  = Y_2+\frac{\lambda_3}{2} v^2, ~~m_{A}^2  = m_{H^\pm}^2+\frac{1}{2}(\lambda_4-\lambda_5) v^2.
 \eeq
The $h$ is the SM-like Higgs, and has no mixing with the inert CP-even Higgs $H$.
Their masses are given as
\beq \label{massh}
 m_{h}^2  = \lambda_1 v^2\equiv (125~{\rm GeV })^2, ~~m_{H}^2  = m_{A}^2+\lambda_5 v^2.
 \eeq

We obtain the masses of fermions via the Yukawa interactions with $\Phi_1$,
 \beq \label{yukawacoupling} - {\cal L} = y_u\overline{Q}_L \,
\tilde{{ \Phi}}_1 \,U_R + y_d\overline{Q}_L\,{\Phi}_1 \, D_R +  y_\ell\overline{L}_L \, {\Phi}_1
\, E_R + \mbox{h.c.}, \eeq
where $\widetilde\Phi_{1}=i\tau_2 \Phi_{1}^*$, $Q_L^T=(u_{Li}\,,d_{Li})$, $L_L^T=(\nu_{Li}\,,\ell_{Li})$ 
with $i$ being generation indices. $U_R$, $D_R$, and $E_R$ are the three generation right-handed fields of
the up-type quark, down-type quark, and charged lepton.
Under the $Z_4$ symmetry, the lepton Yukawa matrix $y_\ell$ to be diagonal. As a result,
the lepton fields ($L_L$, $E_R$) are mass eigenstates.

Under the $Z_4$ symmetry, the $\Phi_2$ is allowed to have $\mu$-$\tau$ interactions \cite{190410908},
\bea\label{lepyukawa2}
- {\cal L}_{LFV} &=&  \sqrt{2}~\rho_{\mu\tau} \,\overline{L^\mu_{L}} \, {\Phi}_2
\,\tau_R  \, + \sqrt{2}~\rho_{\tau\mu}\, \overline{L^\tau_{L}} \, {\Phi}_2
\,\mu_R \, + \, \mbox{h.c.}\,. \eea
The interactions of Eq. (\ref{lepyukawa2}) lead to the $\mu$-$\tau$ LFV couplings of $H$, $A$, and $H^\pm$.
We take the CP-conserving Yukawa matrix, namely that $\rho_{\mu\tau}$ and $\rho_{\tau\mu}$ 
are real and $\rho_{\mu\tau}=\rho_{\tau\mu}\equiv\rho$.

The SM-like Higgs $h$ has the same tree-level couplings to fermions and gauge boson as the SM, and 
has no $\mu$-$\tau$ LFV coupling. The $H$, $A$, and $H^\pm$ have the $\mu$-$\tau$-philic Yukawa couplings
and no other Yukawa couplings. There are no cubic interactions with $ZZ, ~WW$ for the neutral Higgses $A$ and $H$.

\section{Muon $g-2$, LUF in $\tau$ decays, LHC data, and relevant constraints}\label{constraints}
In our calculations, the input parameters are $\lambda_2$, $\lambda_3$, $m_h$, $m_H$, $m_A$ and $m_{H^\pm}$ which
can determine the values of $\lambda_1$, $\lambda_5$ and $\lambda_4$ from Eqs. (\ref{masshp}, \ref{massh}).
  We take $m_h=125$ GeV, and scan over several key parameters in the following ranges:
\bea
&&0 < \rho <1.0,~5~{\rm GeV}<m_H<115~{\rm GeV},\nonumber\\
&&~130~{\rm GeV} < m_A < 900~ {\rm GeV},~90~ {\rm GeV}<m_{H^{\pm}} < 900~ {\rm GeV}.
\label{range}
\eea
The $\lambda_2$ controls the quartic couplings of extra Higgses, and does not affect the observables considered in our paper.
We choose $\lambda_3=\lambda_4+\lambda_5$ which leads the $hHH$ coupling to be absent.
At the tree-level, the SM-like Higgs $h$ has the same couplings to the SM particles as the SM and no exotic decay mode.
Since the extra Higgses have no couplings to quarks, we may safely neglect the bounds of meson observable 
 $\textsf{HiggsBounds}$ \cite{hb1} is used to implement the exclusion
constraints from the searches for the neutral and charged Higgs at the LEP 
at 95\% confidence level. In addition, we consider other observables and constraints:
\begin{itemize}
\item[(1)] Theoretical constraints and the oblique parameters. We use the $\textsf{2HDMC}$ \cite{2hc-1}
 to implement the theoretical
constraints from the vacuum stability, unitarity and
coupling-constant perturbativity, and calculate 
the oblique parameters ($S$, $T$, $U$). According to the recent fit results of Ref. \cite{pdg2018}, we take the following 
values of $S$, $T$, $U$,
\beq
S=0.02\pm 0.10, ~~T=0.07\pm 0.12,~~U=0.00 \pm 0.09. 
\eeq
The correlation coefficients are given by
\beq
\rho_{ST} = 0.92, ~~\rho_{SU} = -0.66, ~~\rho_{TU} = -0.86.
\eeq
The oblique parameters favor one of $H$ and $A$ to have a small mass splitting from $H^\pm$.

\item[(2)] Muon $g-2$. In the model, the new contribution to muon $g-2$ comes from the one-loop diagrams involving 
the $\mu$-$\tau$ LFV coupling of $H$ and $A$ \cite{10010434}, 
\bea
  \delta a_{\mu}=\frac{m_\mu m_\tau \rho^2}{8\pi^2}
  \left[\frac{(\log\frac{m_H^2}{m_\tau^2}-\frac{3}{2})}{m_H^2}
  -\frac{\log(\frac{m_A^2}{m_\tau^2}-\frac{3}{2})}{m_A^2}
\right].
  \label{mua1}
\eea
The Eq. (\ref{mua1}) shows that the new contributions are positive for $m_A>m_H$. This is reason why
we take $m_A > m_H$ in our calculations.

\item[(3)] Lepton universality in the $\tau$ decays.
The HFAG collaboration reported three ratios from pure leptonic processes, and two ratios
from semi-hadronic processes, $\tau \to \pi/K \nu$ and $\pi/K \to \mu \nu$ \cite{tauexp}
\begin{eqnarray} \label{hfag-data}
&&
\left( g_\tau \over g_\mu \right) =1.0011 \pm 0.0015,~\left( g_\tau \over g_e \right)  = 1.0029 \pm 0.0015,\nonumber\\
&&\left( g_\mu \over g_e \right) = 1.0018 \pm 0.0014,~\left( g_\tau \over g_\mu \right)_\pi = 0.9963 \pm 0.0027,\nonumber\\
&&\left( g_\tau \over g_\mu \right)_K = 0.9858 \pm 0.0071,
\end{eqnarray}
with the definitions
\begin{eqnarray} 
&&
\left( g_\tau \over g_\mu \right)^2 \equiv \bar{\Gamma}(\tau\to e
\nu\bar{\nu})/\bar{\Gamma}(\mu\to e \nu\bar{\nu}),\nonumber\\
&&\left( g_\tau \over g_e \right)^2  \equiv \bar{\Gamma}(\tau\to \mu
\nu\bar{\nu})/\bar{\Gamma}(\mu\to e \nu\bar{\nu}),\nonumber\\
&&\left( g_\mu \over g_e \right)^2 \equiv \bar{\Gamma}(\tau\to \mu
\nu\bar{\nu})/\bar{\Gamma}(\tau\to e \nu\bar{\nu}).
\end{eqnarray}
The correlation matrix for the above five observables is 
\begin{equation} \label{hfag-corr}
\left(
\begin{array}{ccccc}
1 & +0.53 & -0.49 & +0.24 & +0.12 \\
+0.53  & 1     &  + 0.48 & +0.26    & +0.10 \\
-0.49  & +0.48  & 1       &   +0.02 & -0.02 \\
+0.24  & +0.26  & +0.02  &     1    &     +0.05 \\
+0.12  & +0.10  & -0.02  &  +0.05  &   1 
\end{array} \right).
\end{equation}

The $\bar{\Gamma}$ denotes the partial width
normalized to its SM value. In this model, we have
\begin{eqnarray} \label{tau-loop}
&&\bar{\Gamma}(\tau\to \mu \nu\bar{\nu})= (1+\delta_{\rm loop}^\tau)^2~(1+\delta_{\rm loop}^\mu)^2+\delta_{\rm tree},\nonumber\\
&&\bar{\Gamma}(\tau\to e \nu\bar{\nu})= (1+\delta_{\rm loop}^\tau)^2,\nonumber\\
&&\bar{\Gamma}(\mu\to e \nu\bar{\nu})= (1+\delta_{\rm loop}^\mu)^2.
\end{eqnarray}
Where $\delta_{\rm tree}$ can give a positive correction to $\tau\to \mu \nu\bar{\nu}$, and is from the tree-level diagram
mediated by the charged Higgs,
\beq
\delta_{\rm tree}=4\frac{m_W^4\rho^4}{g^4 m_{H^{\pm}}^4}.
\label{delta-tree}
\eeq
$\delta_{\rm loop}^\tau$ and $\delta_{\rm loop}^\mu$ are the corrections to vertices $W\bar{\nu_{\tau}}\tau$ and
$W\bar{\nu_{\mu}}\mu$, which are from the one-loop diagrams involving $H$, $A$, and $H^\pm$. Since we
take $\rho_{\mu\tau}=\rho_{\tau\mu}$ in the lepton Yukawa matrix, and therefore $\delta_{\rm loop}^\tau=\delta_{\rm loop}^\mu$. Following 
the results of \cite{tavv-1,tavv-2,190410908},
\beq
\delta_{\rm loop}^\tau=\delta_{\rm loop}^\mu={1 \over 16 \pi^2} {\rho^2} 
\left[1 + {1\over4} \left( H(x_A) +  H(x_H) \right)
\right]\,, 
\label{delta-loop}
\eeq
where $H(x_\phi) \equiv \ln(x_\phi) (1+x_\phi)/(1-x_\phi)$ with $x_\phi=m_\phi^2/m_{H^{\pm}}^2$.

In the model, 
\beq\left( g_\tau \over g_\mu \right)_\pi=\left( g_\tau \over g_\mu \right)_K =\left( g_\tau \over g_\mu \right).
\eeq
We perform $\chi^2_\tau$ calculation for the five observables. The covariance matrix constructed from the data of Eq. (\ref{hfag-data})
and Eq. (\ref{hfag-corr}) has a vanishing eigenvalue, and the corresponding degree is removed in our calculation.
In our discussions we require the value of $\chi^2_\tau$ to be smaller than the SM value, $12.3$.

\item[(4)] Lepton universality in the $Z$ decays. The experimental values of the ratios of the leptonic $Z$ decay
branching fractions are given as \cite{zexp}:
\begin{eqnarray} \label{lu-zdecay}
{\Gamma_{Z\to \tau^+ \tau^- }\over \Gamma_{Z\to e^+ e^- }} &=& 1.0019 \pm 0.0032,\\
{\Gamma_{Z\to \mu^+ \mu^-}\over \Gamma_{Z\to e^+ e^- }} &=& 1.0009 \pm 0.0028,
\end{eqnarray}
with a correlation of $+0.63$. 
In the model, the new contributions to the widths of $Z\to \tau^+\tau^-$ and $Z\to \mu^+\mu^-$ are
from the one-loop diagrams involving the extra Higgs bosons. The ratio of Eq. (\ref{lu-zdecay}) is given
 as \cite{tavv-1,tavv-2,190410908}
\beq 
{\Gamma_{Z\to \tau^+ \tau^- }\over \Gamma_{Z\to e^+ e^- }} 
\approx 1.0+ {2 g_L^e{\rm Re}(\delta g^{\rm loop}_L)+ 2 g_R^e{\rm Re}(\delta g^{\rm loop}_R) \over {g_L^e}^2 + {g_R^e}^2 }\,.
\,
\eeq
with the SM value $g_L^e=-0.27$ and $g_R^e=0.23$. The one-loop corrections $\delta g^{\rm loop}_L$ and $\delta g^{\rm loop}_R$
are from  
\begin{eqnarray} \label{dgLR}
\delta g^{\rm loop}_L &=& {1\over 16\pi^2} \rho^2 \,
\bigg\{
 -{1\over2} B_Z(r_A)- {1\over2} B_Z(r_H) -2 C_Z(r_A, r_H)
   \nonumber \\
&&  + s_W^2 \left[ B_Z(r_A) + B_Z(r_H) + \tilde C_Z(r_A) + \tilde C_Z(r_H) \right] \bigg\} 
\,,\\ 
\delta g^{\rm loop}_R &=& {1\over 16\pi^2} \rho^2 \,
\bigg\{ 2 C_Z(r_A, r_H) - 2 C_Z(r_{H^\pm}, r_{H^\pm}) 
+ \tilde C_Z(r_{H^\pm}) 
\nonumber \\
&& - {1\over2} \tilde C_Z(r_A)  - {1\over2} \tilde C_Z(r_H)      
 +  s_W^2 \left[ B_Z(r_A) + B_Z(r_H) + 2 B_Z(r_{H^\pm})\right.
\nonumber \\
&& \left. +\tilde C_Z(r_A) + \tilde C_Z(r_H) + 4C_Z(r_{H^\pm},r_{H^\pm}) \right] \bigg\} 
\,,
\end{eqnarray}
where $r_\phi = m_\phi^2/m_Z^2$ with $\phi=A,H, H^\pm$, and 
\begin{eqnarray} 
\label{loopftn}
B_Z(r) &=& -{\Delta_\epsilon \over 2} -{1\over4} + {1\over2} \log(r) \,,\\ 
C_Z(r_1,r_2) &=& {\Delta_\epsilon \over4} -{1\over2} \int^1_0 d x \int^x_0 d y\,
\log[ r_2 (1-x) + (r_1 -1) y + x y] \,,\\ 
\tilde C_Z(r) &=& {\Delta_\epsilon \over2}+{1\over2}  - r\big[1+\log(r) \big]
+r^2 \big[ \log(r) \log(1+r^{-1}) \nonumber\\
&&
-{\rm Li_2}(-r^{-1}) \big] -{i \pi\over2}
\left[ 
	1 - 2r + 2r^{2}\log(1+r^{-1})
\right].
\end{eqnarray}
Besides, $\Gamma_{Z\to \mu^+ \mu^-}$ equals to $\Gamma_{Z\to \tau^+ \tau^- }$ for $\rho_{\mu\tau}=\rho_{\tau\mu}$.

(5) The multi-lepton searches at the LHC. The $H$, $A$, and $H^{\pm}$ are mainly produced at the LHC via
the electroweak processes:
\begin{align}
pp\to & W^{\pm *} \to H^\pm A, \label{process1}\\
pp\to &       Z^* \to HA, \label{process2}\\
pp\to & W^{\pm *} \to H^\pm H, \label{process3}\\
pp\to & Z^*/\gamma^* \to H^+H^-. \label{process4}\\
pp\to & Z \to \tau^\pm\mu^\mp H.
\end{align}

When $H$, $A$, and $H^{\pm}$ have small splitting mass, the main decay modes of these Higgses are
\beq
H\to \tau^{\pm}\mu^{\mp},~~~A\to \tau^{\pm}\mu^{\mp},~~~H^\pm\to \tau^\pm \nu_\mu, \mu^\pm \nu_\tau.
\label{decay-taumu}
\eeq
For $m_A~(m_{H^{\pm}}) > m_H +m_Z$,  the following exotic decay modes will open, 
\beq
A\to HZ, ~~~~H^\pm \to H W^\pm. 
\eeq
We perform simulations for the processes employing \texttt{MG5\_aMC-2.4.3}~\cite{Alwall:2014hca} 
with \texttt{PYTHIA6}~\cite{Torrielli:2010aw} and 
\texttt{Delphes-3.2.0}~\cite{deFavereau:2013fsa}, and  impose the constraints 
from all the analysis at the 13 TeV LHC in the latest \texttt{CheckMATE 2.0.28}~\cite{Dercks:2016npn}
and \texttt{Fastjet}~\cite{Cacciari:2011ma}. The analysis we implemented in our previous works~\cite{1908.09845,Pozzo:2018anw} are also included. Besides, we impose the recently published analyses of searching for events with three or more leptons, with up to two hadronical $\tau$ leptons, using 13~TeV LHC 137~fb$^{-1}$ data~\cite{CMS:2021bra}. It improves significantly the limits on new physical particles that decay to leptons. The signal regions of \texttt{4lI}, \texttt{4lJ} and \texttt{4lK}, which require 4 leptons with one or two hadronical $\tau$ leptons in the final states, are most sensitive to our samples, because of the dominated multi-lepton final states in our model.
\end{itemize}

\section{Results and discussions}
\begin{figure}[tb]
  \epsfig{file=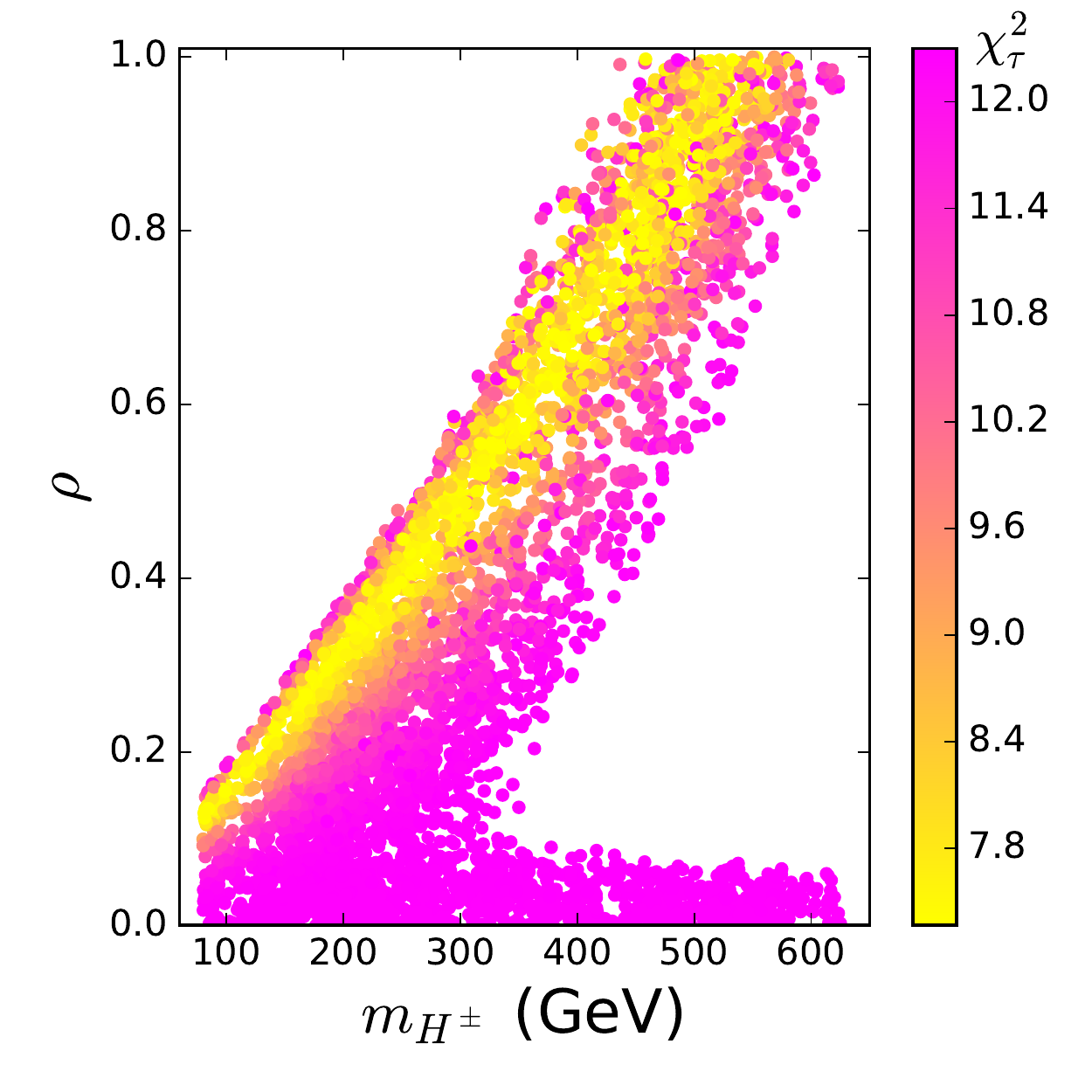,height=8cm}
\vspace{-0.5cm} \caption{The surviving samples satisfying the constraints of "pre-muon $g-2$" and $\chi^2_\tau<$ 12.3
projected on the plane of $\rho$ versus $m_{H^\pm}$.}
\label{taub}
\end{figure}

We impose the constraints of "pre-muon $g-2$" (denoting the theory, 
the oblique parameters, the exclusion limits from the searches for Higgs at LEP), and show the surviving samples 
with $\chi^2_\tau<$ 12.3 fitting the data of LUF in $\tau$ decays in Fig. \ref{taub}.
For a very small $\rho$, the new contributions to $\tau$ decays disappear, and therefore the value of $\chi^2_\tau$ approaches to
the SM value, 12.3. The discrepancy of LUF in $\tau$ decays can be alleviated by enhancing $\Gamma(\tau\to \mu \nu\bar{\nu})$.
From Eq. (\ref{tau-loop}), we can find that $\tau\to \mu\nu\nu$ receives the corrections from the one-loop diagram and tree-level 
diagram mediated by the charged Higgs.
According to Eq. (\ref{delta-loop}), the former tends to give the negative corrections and enhance the value of $\chi^2_\tau$. 
According to Eq. (\ref{delta-tree}), the latter gives the 
positive corrections and reduce the value of $\chi^2_\tau$. In order to obtain $\chi^2_\tau<$ 12.3 for a large $m_{H^\pm}$,
a large $\rho$ is required to make the contributions of tree-level diagram to overcome those of one-loop diagram 
 since the contributions of tree-level diagram are suppressed by $m_{H^\pm}$. For $\chi^2_\tau<$ 9.7,  $\rho$ is always required to increase
with $m_{H^\pm}$ and be larger than 0.11.

\begin{figure}[tb]
  \epsfig{file=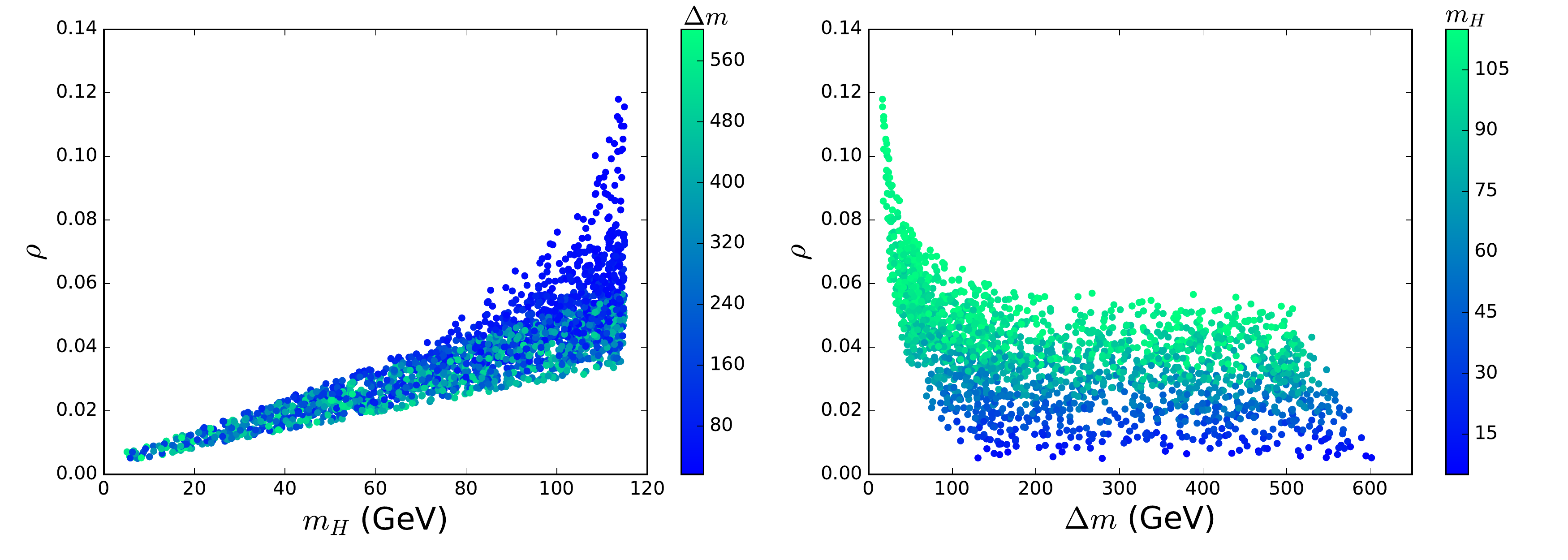,height=6cm}
\vspace{-1.0cm} \caption{The surviving samples satisfying the constraints of "pre-muon $g-2$" and muon $g-2$ anomaly 
projected on the planes of $\rho$ versus $m_{H}$ and $\rho$ versus $\Delta m$.}
\label{g2b}
\end{figure}
\begin{figure}[tb]
  \epsfig{file=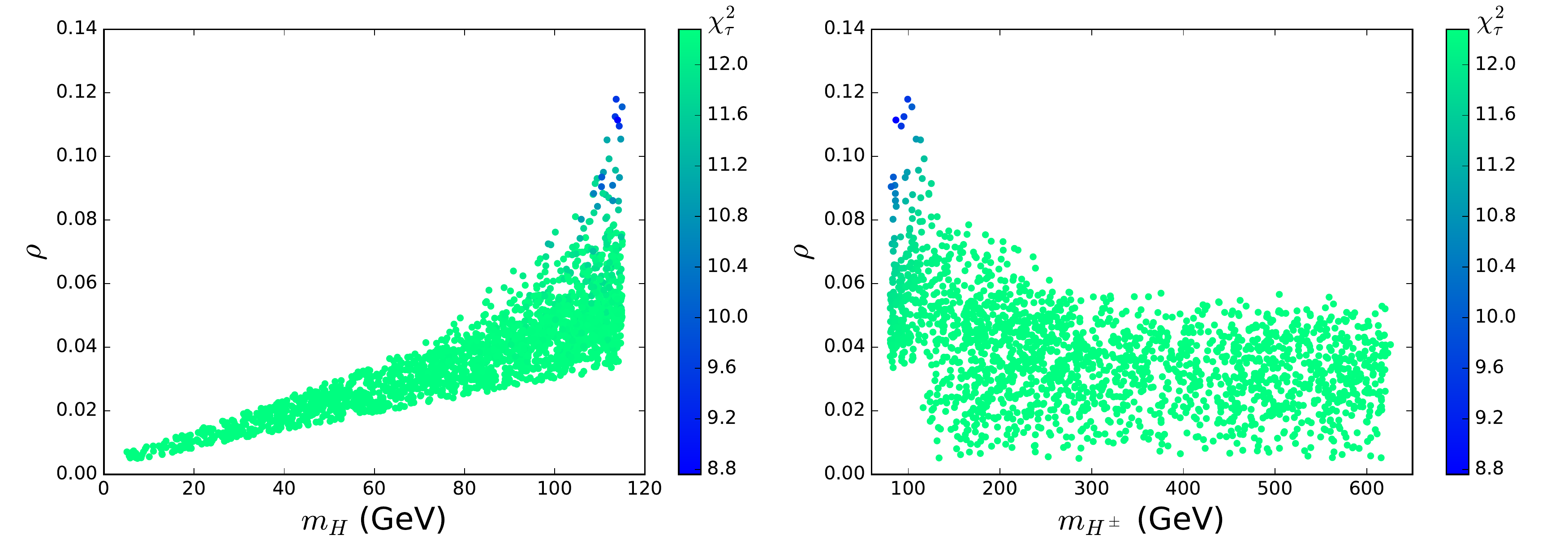,height=6cm}
\vspace{-1.0cm} \caption{The surviving samples satisfying the constraints of "pre-muon $g-2$", muon $g-2$ anomaly,
$\chi^2_\tau<$12.3, and $Z$ decays.}
\label{g2tazb}
\end{figure}
\begin{figure}[tb]
  \epsfig{file=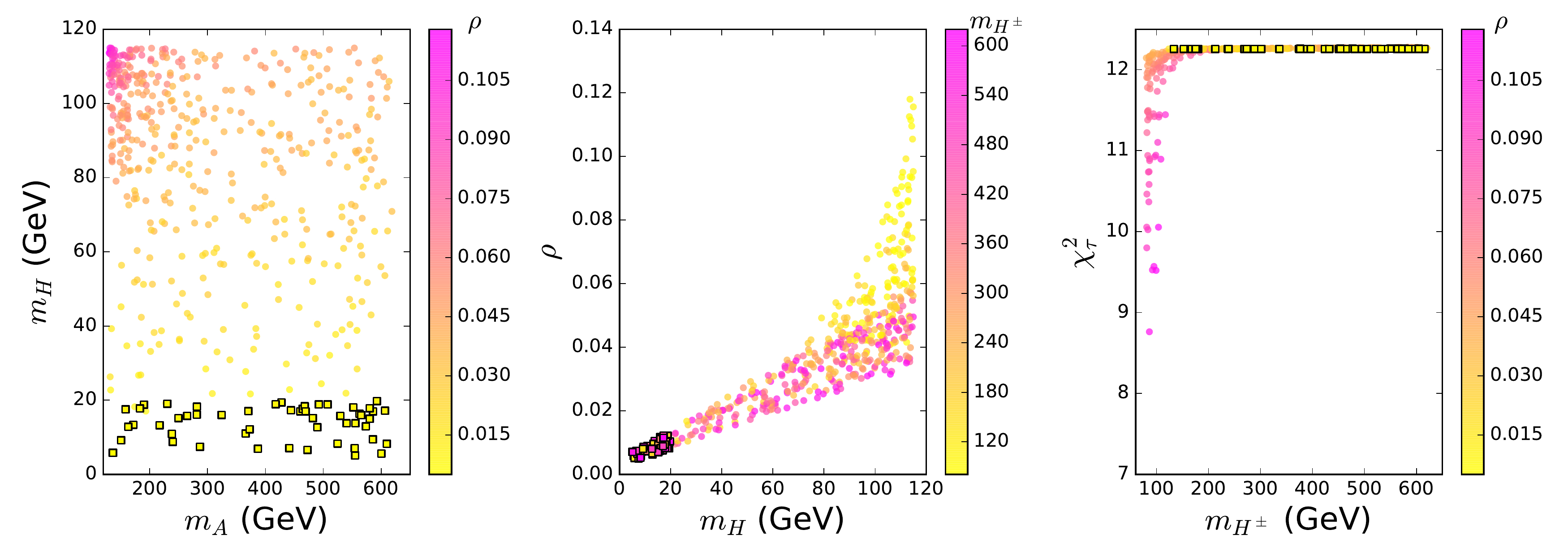,height=6cm}
\vspace{-1.0cm} \caption{The surviving samples on the planes
of $m_H$ versus $m_A$, $\rho$ versus $m_H$, and $\chi^2_\tau$ versus $m_{H^\pm}$. All the samples satisfy the constraints of "pre-muon $g-2$",
 muon $g-2$ anomaly, $\chi^2_\tau<$ 12.3, and $Z$ decays. 
The bullets and squares are excluded and allowed by the direct searches at the LHC.}
\label{lhcb}
\end{figure}

After imposing the constraints of "pre-muon $g-2$" and muon $g-2$, we project the surviving samples on the planes of 
$\rho$ versus $m_H$ and $\rho$ versus $\Delta m$ ($\Delta m=m_A-m_H$) in
Fig. \ref{g2b}. From the Eq. (\ref{mua1}), we can
find that $\Delta a_\mu$ receives a positive correction from
the diagrams involving $H$ and a negative correction from ones involving $A$. As a result, $\Delta a_\mu$ is sizable enhanced 
by a large mass splitting between $m_A$ and $m_H$ ($\Delta m$), and favors $\rho$ to decrease with an increase of $\Delta m$,
as shown in the right panel of Fig. \ref{g2b}.
In addition, the Eq. (\ref{mua1}) shows that the contributions of the diagrams involving $H$ and $A$ to $\Delta a_\mu$ are respectively
suppressed by $m_H^2$ and $m_A^2$. Therefore,  $\Delta a_\mu$ favors $\rho$ to increase with $m_H$, as shown in the
left panel of Fig. \ref{g2b}. The muon $g-2$ anomaly can be explained in the paramete space of $0.005<\rho<0.12$ and 
5 GeV $<m_H<$ 115 GeV.

In Fig. \ref{g2tazb},  we show the surviving samples after imposing the constraints of "pre-muon $g-2$", muon $g-2$, $\chi^2_\tau<$ 12.3, and 
$Z$ decays. Since the muon $g-2$ anomaly favors $0.005<\rho<0.12$, most of the parameter space satisfying $\chi_\tau^2<$ 12.3 
are excluded. 
 From Fig. \ref{g2tazb}, we find that a small $\chi^2_\tau$ favors a large $m_H$ and a small $m_{H^\pm}$.
Since the muon $g-2$ anomaly favors a large $\rho$ for a large $m_{H}$, and such large $\rho$ can enhance the width of 
$\tau\to \mu\nu\nu$ and reduce the value of $\chi^2_\tau$.

After imposing the constraints of the direct searches at the LHC, the surviving 
samples of Fig. \ref{g2tazb} are projected on Fig. \ref{lhcb}.
We find that the direct searches at the LHC impose a stringent upper bound on $m_H$, $m_H<20$ GeV, and
allow 130 GeV $<m_A~(m_{H^\pm})<$ 610 GeV. For a light $H$, the $\tau\mu$ from $H$ 
 decays become too soft to be distinguished at detector, while the $\tau\mu$ from $H$ in $A/H^\pm$ decays are collinear 
because of the large mass splitting between $H$ and $A/H^\pm$. In addition, the $A/H^\pm \to H Z/W^\pm$ decay modes dominate 
the $A/H^\pm$ decays in the low $m_{H}$ region. Thus, in the region of $m_{H}<20$ GeV, the acceptance of above signal region for final state of 
 collinear $\tau\mu$ + $Z/W$ boson quickly decreases. 
For 5 GeV $<m_H<$ 20 GeV, the muon $g-2$ anomaly favors 0.005 $<\rho<$ 0.014.
As a result, the new contributions to the $\tau$ decays are very small, and the $\chi^2_\tau$ approaches to
the value of SM, 12.3.

\section{Conclusion}
In the 2HDM with an abelian discrete $Z_4$ symmetry, one Higgs doublet has the same interactions with fermions as the SM, 
and another inert Higgs doublet only 
has the $\mu$-$\tau$ LFV interactions. After imposing various relevant theoretical and experimental 
constraints, especially for the multi-letpon search at the LHC, we found that the model can explain the muon $g-2$ anomaly within $2\sigma$ range 
in the region of 5 GeV $<m_H<20$ GeV, 130 GeV $< m_A~(m_{H^\pm})<$ 610 GeV, and 0.005 $<\rho<$ 0.014. Meanwhile,
the $\chi^2_\tau$ fitting the data of LFU in the $\tau$ decays
approaches to the SM prediction.

\section*{Acknowledgment}
This work was by the National Natural Science Foundation
of China under grant 11975013, by the Natural Science Foundation of
Shandong province (ZR2017JL002 and ZR2017MA004), and by the Project of Shandong Province Higher Educational Science and
Technology Program under Grants No. 2019KJJ007.

\end{document}